\documentclass[aps,pre,showpacs,twocolumn]{revtex4-1} 	

\usepackage{graphicx, amsmath, amssymb}

\def \lvec{(\kern-.26em(}

\begin{document}

\title{Active regeneration unites high- and low-temperature features in cooperative self-assembly}
\author{Robert Marsland III}
\altaffiliation{Current address: Department of Physics, Boston University, 590 Commonwealth Avenue, Boston, MA 02215}
\email[Email: ]{marsland@bu.edu}
\author{Jeremy England}
\affiliation{Physics of Living Systems Group, Massachusetts Institute of Technology, 400 Technology Square, Cambridge, MA 02139}
\date{\today}
\begin{abstract}
Cytoskeletal filaments are capable of self-assembly in the absence of externally supplied chemical energy, but the rapid turnover rates essential for their biological function require a constant flux of ATP or GTP hydrolysis. The same is true for two-dimensional protein assemblies employed in the formation of vesicles from cellular membranes, which rely on ATP-hydrolyzing enzymes to rapidly disassemble upon completion of the process. Recent observations suggest that the nucleolus, p granules and other three-dimensional membraneless organelles may also demand dissipation of chemical energy to maintain their fluidity. Cooperative binding plays a crucial role in the dynamics of these higher-dimensional structures, but is absent from classic models of 1-dimensional cytoskeletal assembly. In this Letter, we present a thermodynamically consistent model of actively regeneration with cooperative assembly, and compute the maximum turnover rate and minimum disassembly time as a function of the chemical driving force and the binding energy. We find that these driven structures resemble different equilibrium states above and below the nucleation barrier. In particular, we show that the maximal acceleration under large binding energies unites infinite-temperature local fluctuations with low-temperature nucleation kinetics.
\end{abstract}

\maketitle

\section{Introduction}
Kirschner and Mitchison pointed out in the 1980's that GTP hydrolysis in microtubules is responsible for the amazingly high rate of monomer exchange between filaments and the surrounding solvent \cite{Kirschner1986}. An order-of-magnitude estimate using the measured association rates and binding energies shows that the dissociation rates in thermal equilibrium would be far too slow to support the massive structural rearrangements that take place over the course of the cell cycle. The large difference in chemical potential between the GTP and GDP pools in the cytosol allows the microtubule polymerization reaction to break detailed balance, speeding up the dynamics while maintaining the strength and stiffness demanded by the biological function of these structures.

 The coupling of nucleotide hydrolysis to monomer turnover seen in this particular case is in fact a generic feature shared by a variety of intracellular structures, which rely on similar mechanisms of active regeneration to enable timely responses to biochemical signals without sacrificing mechanical integrity. Recent studies have established the Hsp70 family of chaperone proteins as an all-purpose ATP-powered disassemblase, responsible for the rapid disassembly of disordered aggregates as well as of the protein coats that regulate vesicle formation in eukaryotic cells \cite{Sousa2014,Massol2006,Bocking2011}. Phase-separated intracellular droplets and granules appear to be fluidized by similar mechanisms \cite{Amen2015,Hyman2014}, and even the structure of interphase chromatin seems to be set by the competition between equilibrium phase separation and active disassembly \cite{Nuebler2017}. 

All these processes exhibit high levels of cooperativity: the dissociation rate of a given subunit from the structure depends on the presence or absence of multiple neighboring particles. This cooperativity gives rise to nonlinearities that are not present in simple models of 1D cytoskeletal filaments, which allow monomers to dissociate only at the ends of the filament where they have exactly one neighbor. These nonlinearities can make the equilibrium state of cooperative systems relatively insensitive to changes in temperature, pH or other parameters except for a narrow range of values around a well-defined threshold \cite[Ch. 9]{Nelson2008}. This robustness should allow low levels of active regeneration to accelerate the kinetics without significantly affecting the static properties of the structure. But there is no general rule for determining how much acceleration can be tolerated, or what happens to the structure after this limit is reached. 

Several models have recently been developed that combine high-cooperativity equilibrium dynamics with a nonequilibrium driving force, leading to novel hypotheses about bistability in vesicle coat dynamics \cite{Foret2008} and in the behavior of neurotransmitter protein receptors \cite{Burlakov2012}. But these models are formulated in the limit of an infinite thermodynamic force, where at least one of the reaction steps is completely irreversible, and so they are not suitable for investigating the dependence of the system's properties on thermodynamic drive strength. 

In this work, we therefore present a fully reversible model of the intrinsically cooperative self-assembly of a high-dimensional structure, with active regeneration powered by a finite chemical potential difference $\Delta\mu$ between chemical reservoirs. This model allows us to compute the kinetic acceleration at the disassembly threshold as a function of chemical potential difference and monomer binding energy. We confirm that a small nonequilibrium driving force accelerates the kinetics without dramatically modifying any static properties. But at a critical value of $\Delta\mu$, the spectrum of fluctuations suddenly changes, with a corresponding jump in speed. We show that this novel phase combines the purely entropic local fluctuation spectrum of an infinite-temperature equilibrium state with the nucleation barrier that would be expected at the actual temperature.

\section{Model Formulation}

\begin{figure}
	\centering
	\includegraphics[width=9cm]{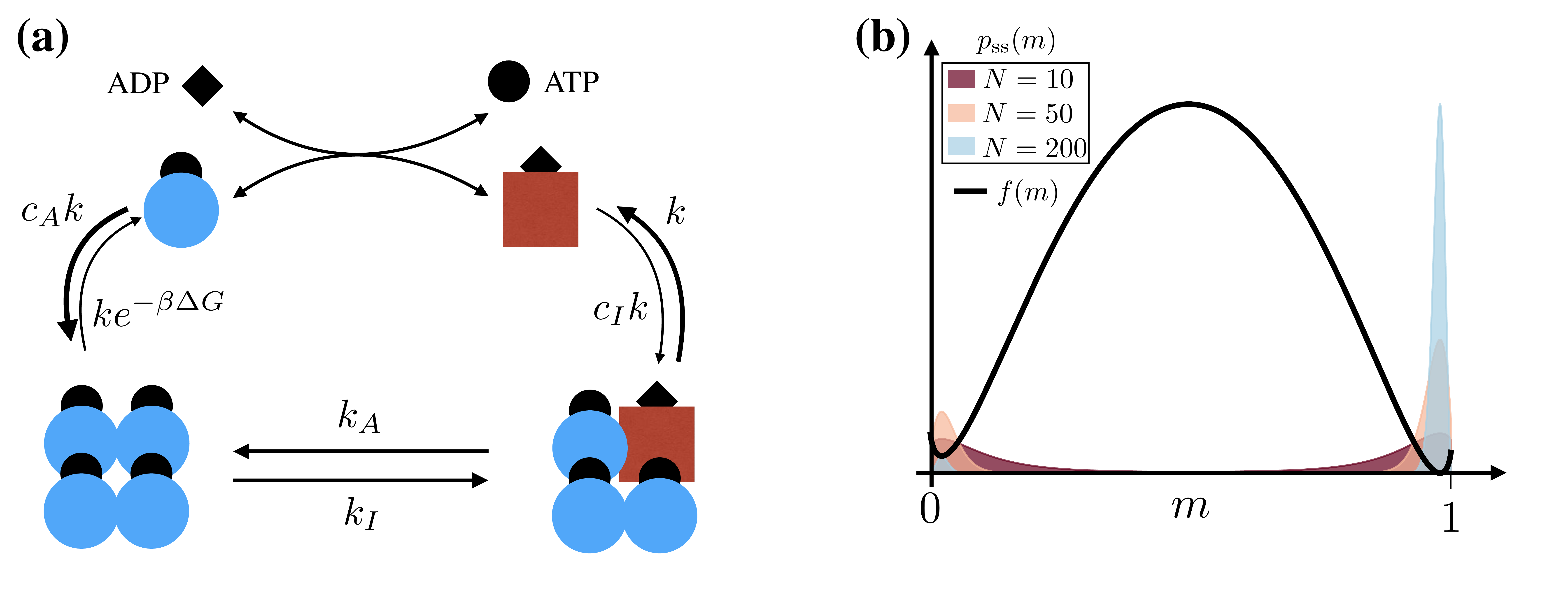}
	\caption{{\bf Themodynamically consistent model of actively regenerating cooperative self-assembly process}. Color online. (a) Blue circles and red squares represent two distinct internal states of the assembling monomers with different binding energies ($\Delta G > 0$ and 0, respectively). Arrows represent binding/unbinding and state-changing reactions. A steady-state probability current is maintained around this cycle by coupling the state change to ATP hydrolysis. The circles are stabilized by binding to ATP (small circles), and the squares by binding to ADP (diamonds). The free energy of hydrolysis biases the state-change reaction of a bound circle towards the non-interacting square state. The square rapidly dissociates into the solution, where nucleotide exchange (double-headed arrows) is much faster than hydrolysis, and transforms the particle back to the circle state. The particle can now bind to the structure again, completing the cycle. (b) Shaded curves are steady-state probability distributions $p_{\rm ss}(m)$ over the occupancy fraction $m$, for increasing values of the system size $N$, with $\Delta\mu = 0$, $\beta J = 8$ and $c_A = 0.02$. Black line is the effective free energy density, defined as $f(m) = - \lim_{N\to\infty} p_{\rm ss}(m)/N$. }
	\label{fig:model_fig}
\end{figure}

Our model describes a solution of proteins that can exist in two different conformational states with identical internal free energy, illustrated by circles and squares in Figure \ref{fig:model_fig}: an ``active'' form that can bind to other active proteins to generate a larger structure, and an ``inactive'' form with no binding ability. Inactive proteins in solution at concentration $c_I$ stochastically enter and leave free binding sites of an existing structure, with Poisson rates $k^{\rm on}_I = c_I k$ and $k^{\rm off}_I = k$ per site, respectively. Since these proteins do not interact with each other or with the active proteins, neither rate is affected by the occupancy of other sites in the structure. For notational simplicity, we have chosen the units of concentration such that $c_I = 1$ is the value at which these non-interacting proteins would occupy half of the available sites. Active proteins, at concentration $c_A$, stick to other active proteins when they enter the structure, and the ratio of their association and dissociation rates is determined by the binding energy $\Delta G$. These proteins bind to available sites at a rate $k^{\rm on}_A = c_A k$ per site, and dissociate at a rate $k^{\rm off}_A = k \exp[-\beta \Delta G]$ where $\beta = 1/(k_B T)$ is the inverse thermal energy scale. The binding energy is proportional to the fraction $m \in [0,1]$ of the $N$ binding sites of the fully assembled structure that are currently occupied, so that $\Delta G = Jm$ for some constant $J$. This form of $\Delta G$ ignores the effect of spatial heterogeneity on the binding kinetics, but is commonly used in statistical physics to construct a qualitatively correct and analytically tractable theory of phase transitions \cite[Ch. 5]{Kardar}.

The existence of two internal states with different binding energies allows for a thermodynamic driving force to accelerate the kinetics. This acceleration is achieved by biasing the active-inactive transition in different directions, depending on whether the protein is in solution or in the structure \cite{Boekhoven2015}. In solution, the transition should be biased towards the active state, to maintain a large pool of proteins ready to assemble. But in the structure, it should be biased towards the inactive state, so that bound proteins can be rapidly ejected and replaced.

This cycle breaks detailed balance, and so is impossible at thermal equilibrium. Cells power these cycles with nucleotide hydrolysis. In actin filaments, for example, the active monomer conformation is stabilized by binding to ATP, and the inactive conformation is stabilized by binding to ADP \cite{Korn1987}. When actin monomers are in solution, exchange of ADP for ATP takes place at a much faster rate than ATP hydrolysis, and the large concentration difference between the nucleotides under physiological conditions biases the transition towards the active ATP-bound state. But when a monomer is in a filament, hydrolysis is much faster than nucleotide exchange, and the large free energy of this reaction reverses the bias.

Figure \ref{fig:model_fig} illustrates how we implemented this general strategy within our model of cooperative assembly. We assume nucleotide is fast enough compared to hydrolysis for proteins in solution that the latter reaction is negligible. The cytosolic concentrations $c_A$ and $c_I$ of the two conformations are thus fixed at the equilibrium values determined by the ATP/ADP ratio. But within the structure, nucleotide exchange is forbidden. The transition from the active to the inactive conformation takes place at a rate $k_I$, and is necessarily coupled to ATP hydrolysis. The reverse process then involves reversing the hydrolysis reaction, with the result that the rate $k_A$ for returning to the active state is much slower than it otherwise would be. Since the two conformations have the same intrinsic free energy, the free energy change during the transition comes only from the altered interactions of the protein with the rest of the structure, and from the nucleotide hydrolysis. Local detailed balance \cite{Seifert2012,Marsland2018} then requires that
\begin{align}
\frac{k_I}{k_A} = e^{\beta(\Delta \mu_0 - \Delta G)}
\end{align}
where $\Delta \mu_0$ is the free energy released in the hydrolysis reaction, related to the full chemical potential difference $\Delta\mu$ by 
\begin{align}
\label{eq:chempot}
\Delta \mu_0 = \Delta \mu - k_B T \ln \frac{\rm [ATP]}{\rm [ADP]} = \Delta \mu - k_B T \ln \frac{c_A}{c_I}.
\end{align}
and the second equality results from rapid nucleotide exchange in solution.

To study kinetic acceleration in this model, we had to identify a fixed timescale to serve as a point of reference. In many biochemical contexts, association rates are mainly determined by the speed of diffusion to the binding site, and are insensitive to changes in binding energy. We therefore chose $1/k$ as the basic timescale, and set $k = 1$ for the remainder of the analysis.

A trivial way to accelerate the kinetics without breaking detailed balance would be to increase $c_I$ or $k_I$, so that most of the particles in the structure are inactive, and the typical dissociation rate increases from $k^{\rm off}_A = \exp[-\beta\Delta G]$ to $k^{\rm off}_I = 1$. But since the inactive particles are non-interacting, this would really represent a transient density fluctuation in a concentrated solution of freely diffusing particles, and not a process of self-assembly. We therefore restricted our attention to the regime $c_I,k_I \ll  1$, so that inactive monomers dissociate from the lattice much faster than they are added from the solvent or created from active monomers in the structure. 

\section{Steady-state solution}
We proceeded to quantify the maximum steady-state turnover rate and the minimum time required for total disassembly under these constraints. To this end we obtained a closed set of dynamical equations for the structure occupancy $m$ by eliminating the short-lived states with bound inactive particles from the original model, as described in the Appendix \cite{Pigolotti2008}. In the resulting coarse-grained model, the occupancy can stochastically increase by an increment $\Delta m = 1/N$ in a Poisson process with rate
\begin{align}
w_+(m) &=  N(1-m)c_A[1+e^{-\beta\Delta \mu}q(m)]
\end{align}
and can decrease by $1/N$ with rate
\begin{align}
\label{eq:wm}
w_-(m) =  Nm e^{-\beta Jm}[1+q(m)]
\end{align}
where
\begin{align}
\label{eq:q}
q(m) \equiv \frac{k_I}{e^{-\beta Jm}+k_I e^{-\beta\Delta \mu_0}}
\end{align}
contains the contribution of transient visits to the inactive state. This contribution grows with increasing $\Delta \mu_0$, as the transition to the inactive state becomes more and more irreversible.  

In the limit of large $N$, the steady-state distribution generated by these dynamics scales as $p_{\rm ss}(m) \propto e^{-Nf(m)}$, and the function $f(m)$ can be calculated analytically, as shown in the Appendix:
\begin{align}
f(m) =& f_{\rm eq}(m) + \frac{1}{\beta J}\bigg({\rm Li}_2\left[-k_I(e^{-\beta\Delta \mu_0} + e^{-\beta\Delta \mu})e^{\beta Jm}\right]\nonumber\\
\label{eq:LDF}
&- {\rm Li}_2\left[-k_I(e^{-\beta\Delta \mu_0} + 1)e^{\beta Jm}\right]\bigg) + C.
\end{align}
The first term is related to the equilibrium free energy for a fully coupled Ising model:
\begin{align}
\label{eq:feq}
f_{\rm eq}(m) =m \ln m + (1-m) \ln (1-m) - m \ln c_A - \frac{\beta J}{2}m^2
\end{align}
and the second is expressed in terms of the dilogarithm function
\begin{align}
{\rm Li}_2(z) \equiv \sum_{k=1}^\infty \frac{z^k}{k^2},
\end{align}
with a normalization constant $C$ added at the end.


The probability $p_{\rm ss}(m)$ converges in the $N\to\infty$ limit to a delta function centered on the occupancy $m^*$ that minimizes $f(m)$, as illustrated in Figure \ref{fig:model_fig}. For $\beta J > 4$, $f(m)$ generically exhibits (at least) two local minima, one at high $m$ corresponding to an assembled phase, and one at low $m$ corresponding to a disassembled phase. 

\begin{figure}
	\centering
	\includegraphics[width=8.5cm]{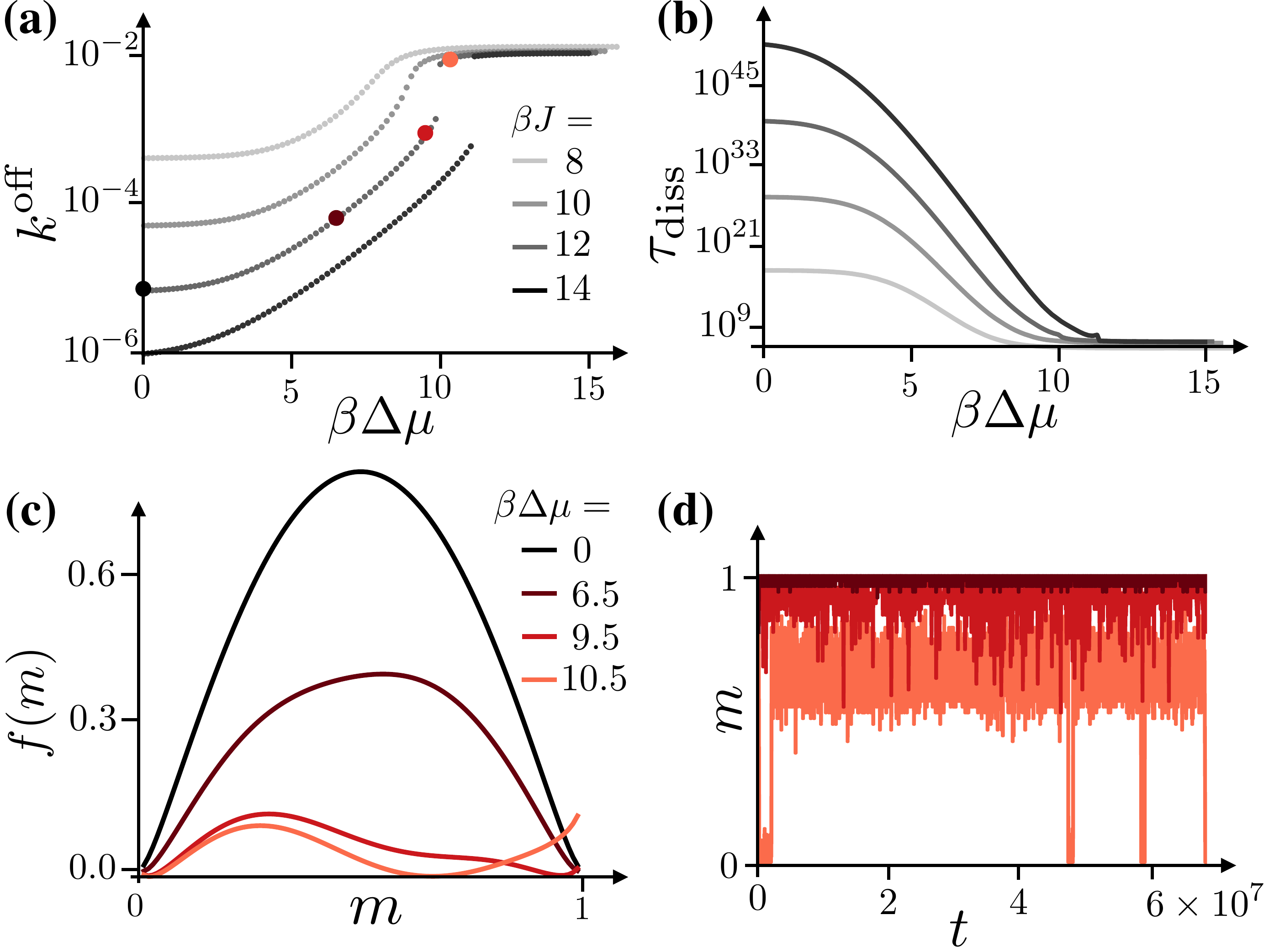}
	\caption{{\bf Active regeneration accelerates turnover and disassembly.} Color online. (a) Local turnover rate $k^{\rm off}$ in assembled phase for four values of $\beta J$. $c_A$ and $\Delta\mu_0$ are tuned to the coexistence point, with $c_I = 0.0001$. (b) Mean time $\tau_{\rm diss}$ for switching from high to low $m$ states under the same conditions, for a structure of size $N = 100$. (c) Effective free energy landscapes $f(m)$ on the $\beta J = 12$ curve at the four values of $\Delta\mu$ indicated by the dots with corresponding colors in panel (a). (d) Stochastic trajectories for $N = 100$ at the three nonzero $\Delta\mu$ values, plotted in the same colors.}
	\label{fig:MFPT}
\end{figure}

\section{Turnover and disassembly speed}
In this limit, we can approximate the steady-state turnover rate per particle by evaluating $w_-$ at the global minimum $m^*$:
\begin{align}
\label{eq:koff}
k^{\rm off} \equiv \frac{w_-(m^*)}{Nm^*} =  e^{-\beta Jm^*}[1+q(m^*)].
\end{align}
We computed the maximum value of $k^{\rm off}$ in the assembled phase as a function of $J$ and $\Delta\mu$, with $k_I = 0.01$ and $c_I = 0.0001$ held fixed to ensure they stay sufficiently small. This maximum occurs when $\Delta\mu_0$ is tuned to the coexistence point, so that any further acceleration would send the system over the threshold into the disassembled phase.

At $\Delta\mu = 0$, $k^{\rm off} \approx 2e^{-\beta J}$ is several orders of magnitude smaller than the rate for free diffusion out of a binding site. At the other extreme, $\Delta\mu \to \infty$, every inactivation reaction ends in ejection from the lattice, so $k^{\rm off} \approx k_I $. Panel (a) of Figure \ref{fig:MFPT} shows how the turnover rate crosses over from the former limit to the latter, for four different values of $\beta J$ spanning a physiologically relevant range of 5 to 8 kcal/mol at $T = $ 300 K. $k^{\rm off}$ increases rapidly for moderate values of $\beta\Delta\mu$, while the qualitative shape of the effective free energy landscape remains the same, with narrow minima near $m=0$ and $m=1$, and a single barrier in between. Near $\Delta \mu = 10 k_B T$, however, the landscape changes drastically, and an additional local minimum emerges. When this local minimum becomes the global minimum, $k^{\rm off}$ discontinuously jumps to its high-$\Delta\mu$ limiting value. This discontinuity corresponds to a first-order phase transition in the $N\to\infty$ limit. Notably, the physiological value of $\Delta\mu \sim 20 k_B T$ for ATP/ADP is more than sufficient to cross this transition where it exists and reach the plateau for all four coupling strengths.

The effect of active regeneration is magnified when we consider the mean first-passage time $\tau_{\rm diss}$ for the global transition from the assembled to the disassembled state, when the system is taken across the phase boundary by an infinitesimal change in one of the parameters \cite{vanKampen}:
\begin{align}
\label{eq:MFPT}
\tau_{\rm diss} &= \sum_{m=m_0}^{m^*} \sum_{m' = m}^1 \frac{p_{\rm ss}(m')}{p_{\rm ss}(m)w_-(m)}\\
\label{eq:MFPTapp}
&\sim e^{N[f(m^\dagger) - f(m^*)]}.
\end{align}
Here $m^\dagger$ is the location of the largest value of $f(m)$ between its two local minima, and the second line represents the qualitative behavior of $\tau_{\rm diss}$ in the limit of large $N$. See the Appendix for derivations of these expressions. The factor of $N$ in the exponent of Equation (\ref{eq:MFPTapp}) can rapidly inflate $\tau_{\rm diss}$ to astronomical values. Panel (b) of Figure \ref{fig:MFPT} shows how $\tau_{\rm diss}$ decreases as a function of $\Delta \mu$ when $N = 100$, for the same four values of $\beta J$ as in panel (a). The disassembly time is mainly determined by the height of the nucleation barrier $f(m^\dagger)-f(m^*)$, which remains unchanged at the nonequilibrium phase transition observed above, and so the jump in $\tau_{\rm diss}$ at the transition point is barely visible on the plot. But the slope of the exponential decrease is extremely large, so that a small change in $\Delta\mu$ can shift the disassembly transition from being effectively impossible to being observable on accessible timescales. This is illustrated in panel (d), where the shift in $\Delta\mu$ from 9.5 to 10.5 $k_BT$ allows several spontaneous assembly-disassembly transitions to take place over the plotted timespan. 

\section{Effective temperatures}
\begin{figure}
	\centering
	\includegraphics[width=8.5cm]{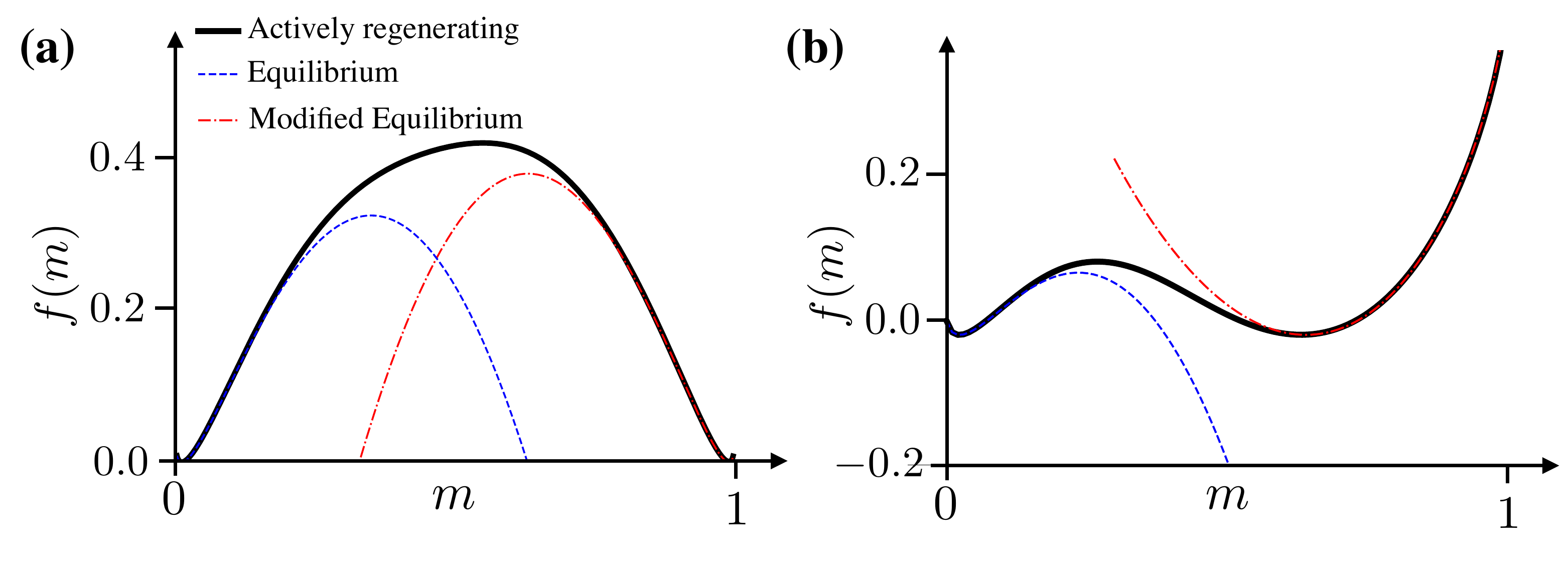}
	\caption{{\bf Driven steady state combines features of two equilibrium landscapes}. Color online. (a) Effective free energy landscape with $\Delta\mu = 6.5 k_B T$, below the nonequilibrium phase transition (black), compared with the equilibrium landscape at the same temperature/coupling $\beta J = 12$ and monomer concentration $c_A = 8.1 \times 10^{-3}$ (blue) as well as a modified equilibrium landscape with the same temperature but a smaller monomer concentration $c_A = 8.8\times 10^{-4}$ (red). (b) Same plots for $\Delta \mu = 15.2 k_B T$, except that the modified equilibrium landscape (red) is obtained by making the temperature infinite, and setting the (constant) off-rate equal to $k_I = 0.01$. }
	\label{fig:newphase}
\end{figure}

Having established that active regeneration can generate substantial kinetic acceleration without disassembling the structure, we proceeded to investigate the extent to which the fluctuations and dynamics of these accelerated states can be captured by an equilibrium model with modified parameters. Figure \ref{fig:newphase} compares the effective free energy of the actively regenerating system on both sides of the nonequilibrium phase transition to similar equilibrium free energy landscapes. The low-$m$ behavior always agrees with $f_{\rm eq}(m)$ at the actual temperature, coupling and monomer concentration. But as $m$ increases, $f(m)$ smoothly transitions to a different equilibrium landscape with altered parameters. When $\Delta\mu$ is below the phase transition, as in panel (a), this altered landscape is obtained by decreasing $c_A$ while keeping the temperature fixed, until the high-$m$ local minimum agrees with the driven system, as shown in the Appendix. This causes a small increase in the turnover rate for the equilibrium model to $k_{\rm off} = 6.7\times 10^{-6}$, almost ten times less than the value of $6.3\times 10^{-5}$ observed in Figure \ref{fig:MFPT}. The barrier to disassembly, however, is significantly lowered by this concentration shift, causing an exponential drop in the disassembly time $\tau_{\rm diss}$. The equilibrium model slightly overestimates the barrier reduction, predicting a $\tau_{\rm diss}$ eight-fold smaller than the actual value in the plotted example.

On the other side of the nonequilibrium phase transition, the high-$m$ part of the landscape is completely different. The matching equilibrium landscape has an effective temperature higher than the critical temperature $T_c = J/4k_B$, and only contains a single local minimum. When $e^{-\beta \Delta \mu_0} \ll e^{-\beta Jm^*}/k_I$, the dynamics near the steady state become independent of $J$, and the effective temperature is infinite, as illustrated in panel (b) of Figure \ref{fig:newphase} and derived in the Appendix. But the actual system still contains the nucleation barrier from the low-$m$ regime, which now determines the disassembly kinetics. 

The emergence of this high effective-temperature phase with a low-temperature nucleation barrier is a very robust phenomenon, which should also arise in more complex models. The essential assumption is that the operation of the regeneration process on a given particle is agnostic to the presence or absence of other particles in the structure. At high enough coupling, active regeneration is the dominant pathway for particle removal in the assembled phase, and at high enough $\Delta \mu_0$, the thermodynamically required dependence of the reverse rate on the coupling becomes irrelevant. In this regime, all binding sites behave independently, and the statistics can be determined by purely entropic considerations. Below the nucleation barrier, on the other hand, active regeneration becomes irrelevant, as long as it is sufficiently slow compared to the spontaneous dissociation rate (which we guaranteed here by requiring $k_I \ll 1$). As we pointed out above, this assumption is necessary if assembly is to proceed at all, and does not limit the generality of the argument. 

\section{Discussion}
We have constructed a minimal model of active regeneration in cooperative self-assembly, and have shown that it can significantly accelerate monomer turnover and assembly/disassembly transitions, while maintaining a sharp distinction between assembled and disassembled states. This confirms that chemically powered regeneration can produce the combination of structural resilience and rapid responsiveness characteristic of living systems \cite{Manu2014}. When the chemical driving force was weak, we observed that the local spectrum of fluctuations was only slightly modified, corresponding to an equilibrium system at the same temperature but lower monomer concentration. In this case, we expect that the effective monomer concentration even in more detailed models may be predictable from fundamental thermodynamic bounds, following a recently developed method for analyzing nonequilibrium assembly processes under moderate drive strength \cite{Nguyen2016}. In the limit of strong driving, we identified a novel nonequilibrium phase with infinite effective temperature, which preserves the nucleation barrier from the original temperature. 

This is the regime we should find when regeneration is driven by nucleotide hydrolysis under physiological conditions, with $\Delta\mu \sim 20 k_B T$. So far, a number of intracellular structures from liquid droplets \cite{Brangwynne2015} to much smaller protein aggregates \cite{Narayanan2017} seem to be well-described by models of equilibrium phase separation, but this is consistent with the predicted persistence of the equilibrium nucleation barrier in the strong driving regime. Testing our predictions will require further investigation of the statistics of the assembled phase. We expect that our analysis will be especially applicable to droplets whose condensation is regulated by the phosphorylation state of the monomers \cite{Li2012}, where a difference in phosphorylation rates within and outside the droplet could set up a dissipative cycle analogous to the one we have described. 

Finally, although we designed our model to capture the physics of high-dimensional structures, our results may also have implications for the behavior of cytoskeletal filaments. Several mechanisms have been proposed that add effective cooperativity to these 1-dimensional structures through length-dependent arrival of specialized molecules that modify the dynamics \cite{Mohapatra2016}. A thermodynamic analysis of these systems raises additional challenges beyond those produced by the intrinsic cooperativity of higher-dimensional models, because the energetics of the underlying molecular motor transport and enzyme activity would also have to be explicitly accounted for. But an exploration of the connections between these models could provide interesting avenues for future work.


\begin{acknowledgments}
RM acknowledges Government support under and awarded by DoD, Air Force Office of Scientific Research, National Defense Science and Engineering Graduate (NDSEG) Fellowship, 32 CFR 168a, and through NIH NIGMS grant 1R35GM119461. We thank Zachary Slepian, Jordan Horowitz, Kirill Korolev and Pankaj Mehta for helpful comments.
\end{acknowledgments}

\bibliographystyle{unsrt}
\bibliography{/Users/robertmarsland/Documents/library}

\section*{Appendix}
This Appendix contains derivations of the following expressions:
\begin{enumerate}
	\item coarse-grained rates $w_+$ and $w_-$, Equations (3)-(5)
	\item effective free energy landscape $f(m)$, Equations (6)-(7)
	\item disassembly time $\tau_{\rm diss}$, Equations (10)-(11).
\end{enumerate}

It also includes additional plots comparing effective free energy landscapes $f(m)$ with equilibrium landscapes, and an analytic derivation of the infinite effective temperature of the novel high-$\Delta\mu$ phase.

\subsection{Coarse-Grained Rates}
Pigolotti \emph{et al.} have described a systematic procedure for eliminating short-lived states from a Markov process, which exactly preserves the stationary state of the original process, and approximates the dynamics to arbitrary precision in the limit of vanishing lifetime of the fast states \cite{Pigolotti2008}. In this section, we give a heuristic motivation for this procedure, and describe how we applied it to our system.

In the main text, we required that the inactive monomers bind so poorly that they make up a negligible percentage of the total structure occupancy at any given time, by imposing
\begin{align}
\label{eq:kcond}
c_I, k_I \ll 1,
\end{align}
in units of time where the dissociation rate of inactive particles is equal to 1. This guarantees that the incoming rates are much smaller than at least one of the outgoing rates, so that the steady state will have much more probability in the states with no particle or a bound active particle in a given site than in states with an inactive particle there. 

Now the state of the system is entirely specified by the occupancy $m$, and there are two ways for a site to change state: either through direct association/dissociation of an active monomer from the solution, or by transiently passing through the inactive form. Since the exit rate from the inactive state is much faster than the time scale of the dynamics of interest (set by the small inactivation rate $k_I$) we can approximate the second pathway as its own Markov process. The rate for dissociation of a given particle via the inactive conformation is simply the product of the inactivation rate $k_I$ and the probability that the site ends up with a different occupancy after the next jump (instead of returning to its starting point). This probability is equal to the ratio of the dissociation rate of the inactive particle (which equals 1 in our units) to the total rate of disappearance of the inactive particle $k_A+1$, which includes the rate of return to the active state without leaving the structure.  Likewise, the association rate through this pathway is the product of the rate $c_I$ for adding an inactive particle to the structure and the probability that it relaxes to the active state instead of returning to the solvent.

Thus the rates for adding and removing active particles via the inactive state are given by
\begin{align}
k^{\rm off}_{AI} &= k_I \frac{1}{1+k_A}\\
k^{\rm on}_{AI} &= c_I \frac{k_A}{1+k_A},
\end{align}
in accordance with Equation (9) of \cite{Pigolotti2008}. 

It is important to note that this procedure preserves the thermodynamics of the original process. The entropy released into the environment in the original model when an active particle is removed via the inactive state is simply the logarithm of the product of the rate ratios (cf. \cite{Seifert2012}):
\begin{align}
\Delta S = k_B \ln \frac{k_I}{k_A}\frac{1}{c_I}
\end{align}
The entropy released in the coarse-grained model is the log-ratio of the new rates, which gives the same value, since the denominators of the fractions cancel out:
\begin{align}
\Delta S = k_B \ln \frac{k^{\rm off}_{AI}}{k^{\rm on}_{AI}} = k_B \ln \frac{k_I \times 1}{k_A \times c_I}.
\end{align}

The total rate $w_+(m)$ for the transition from occupancy $m$ to $m+ 1/N$ is the sum of the rates of all possible ways of accomplishing this transition: particles can be added to any of the $N(1-m)$ free sites, and they can be added to each site by either of the two pathways. Similarly, the $m$ to $m-1/N$ transition rate $w_-(m)$ includes the two removal rates for all $Nm$ particles currently in the structure:
\begin{align}
w_+(m) &= N(1-m)\left( c_A  + c_I  \frac{k_A}{k_A +  1}\right)\\
w_-(m) &= Nm \left(e^{-\beta Jm} + k_I \frac{1}{k_A + 1}\right),
\end{align}
We can now to express both rates in terms of the thermodynamic quantities $\Delta\mu$ and $\Delta \mu_0$, using Equation (2) from the main text to obtain
\begin{align}
\label{eq:Chem}
e^{\beta\Delta \mu} &= \frac{c_A}{c_I} e^{\beta \Delta \mu_0}\\
\label{eq:Free}
&= \frac{c_A k_I}{c_I k_A} e^{\beta Jm}.
\end{align}

Substituting for $c_I$ and $k_A$ in favor of $\Delta\mu$ and $\Delta \mu_0$, we have:
\begin{align}
\label{eq:wp}
w_+(m) &=  N(1-m)c_A[1+e^{-\beta\Delta \mu}q(m)]\\
w_-(m) &=  Nm e^{-\beta Jm}[1+q(m)]
\end{align}
where
\begin{align}
q(m) \equiv \frac{k_I}{e^{-\beta Jm}+k_I e^{-\beta\Delta \mu_0}}.
\end{align}
These are the expressions given in Equations (3)-(5) of the main text.

\subsection{Effective Free Energy Landscape}
\label{sec:free}
The coarse-grained dynamics are one-dimensional, with hard boundaries at $m=0$ and $m=1$, so they cannot support any steady currents. The average rate of jumps from $m$ to $m+1/N$ values has to equal the average rate of jumps from $m+1/N$ to $m$ in order to keep the probability distribution $p(m)$ stationary:
\begin{align}
\label{eq:detailedw}
w_+(m) p_{\rm ss}(m) = w_-(m+1/N) p_{\rm ss}(m+1/N).
\end{align} 
This detailed balance relation implies that the model could also describe an undriven system whose free energy landscape is given by the logarithm of $p_{\rm ss}(m)$. But the functional dependence of these energies on $m$ does not resemble any readily identifiable physical situation.

To analyze the phase behavior of the model, we are interested in the thermodynamic limit where $N\to\infty$. For this reason, we discussed the steady state distribution in terms of the effective free energy density $f(m)$, defined by
\begin{align}
f(m) \equiv -\lim_{N\to\infty} \frac{1}{N}\ln p_{\rm ss}(m).
\end{align}
The derivative of this quantity is related to the ratio of rates via Equation (\ref{eq:detailedw}):
\begin{align}
\frac{df}{dm} &\equiv \lim_{N\to\infty} \frac{f(m+1/N) - f(m)}{1/N}\\
&= \lim_{N\to\infty} \ln \frac{p_{\rm ss}(m)}{p_{\rm ss}(m+1/N)}\\
\label{eq:dfdm}
&= \lim_{N\to\infty} \ln \frac{w_-(m+1/N)}{w_+(m)}\\
&= \ln \frac{w_-(m)}{w_+(m)}
\end{align}
If we substitute in the expressions for $w_-$ and $w_+$ obtained in the previous section, we find
\begin{align}
\frac{df}{dm}&= \ln m - \ln (1-m) - \beta Jm - \ln c_A + \ln \frac{1 +  q}{1 +  e^{-\beta \Delta\mu} q}.
\end{align}

We can now integrate both sides of this equation to find $f(m)$ up to a constant of integration, which will be used to normalize the probability distribution $p_{\rm ss}(m)$. The integrals of the first four terms are straightforward, so that the equilibrium free energy density at $\Delta \mu = 0$ takes on the familiar form: 
\begin{align}
f_{\rm eq}(m) =  m \ln m + (1-m) \ln (1-m) - m \ln c_A  - \frac{\beta J}{2}m^2
\end{align}
as stated in Equation (7) of the main text.

The integral of the final term has no elementary expression, but can be evaluated in terms of the dilogarithm function
\begin{align}
{\rm Li}_2(x) \equiv  \sum_{k=1}^\infty \frac{z^k}{k^2} = -\int dx\, \frac{\ln(1-x)}{x}
\end{align}
to give
\begin{align}
f(m) =& f_{\rm eq}(m) + \frac{1}{\beta J}\bigg({\rm Li}_2\left[-k_I(e^{-\beta\Delta \mu_0} + e^{-\beta\Delta \mu})e^{\beta Jm}\right]\nonumber\\
&- {\rm Li}_2\left[-k_I(e^{-\beta\Delta \mu_0} + 1)e^{\beta Jm}\right]\bigg) + C.
\end{align}
where the constant of integration $C$ can be used to normalize the distribution. This is Equation (6) of the main text.

\subsection{Disassembly Time}

The cooperativity of this model generates a distinct threshold for disassembly, which becomes a sharp first-order phase transition as $N\to\infty$. Close to the threshold, a small change in any of the parameters can shift the probability of finding the system in the assembled phase from nearly 1 to nearly 0. An important kinetic property of the system is the rate at which it switches to the disassembled phase after this parameter change. We quantified this rate using the mean first-passage time $\tau_{\rm diss}$ for the system to reach the low-$m$ local minimum $m_0$ of $f(m)$, starting from the high-$m$ local minimum $m^*$. Since the dynamics are continuous across the transition, we can compute this quantity at the coexistence point where both these local minima have the same value of $f(m)$, and use this as a good approximation for the disassembly time. 

We can derive an expression for $\tau_{\rm diss}$ in terms of the steady-state distribution $p_{\rm ss}(m)$, following \cite[XII.2]{vanKampen}. The argument starts from a self-consistency equation for the mean first-passage time $\tau_{m,m_0}$ from an arbitrary $m$ to the disassembled state $m_0$. We initialize the system at $m$, then wait for a short amount of time $\delta t$. Now the system could be at $m+\Delta m$ with probability $w_+(m)\delta t$, at $m - \Delta m$ with probability $w_-(m)\delta t$, or remain at $m$ with probability $1 - [w_+(m) + w_-(m)]\delta t$ (where $\Delta m = 1/N$). The probability of going more than one step away from $m$ has a probability proportional to $\delta t^2$. Since we have waited at time $\delta t$, the mean time to reach $m_0$ is now equal to $\tau_{m,m_0} - \delta t$. The new waiting time can also be calculated by averaging the mean first-passage times of the possible current system states, weighted by their probabilities. These two ways of calculating the average remaining time need to give the same answer, yielding:
\begin{align}
\tau_{m,m_0} -\delta t =& \tau_{m+\Delta m,m_0} w_+(m)\delta t +  \tau_{m-\Delta m,m_0} w_-(m)\delta t \nonumber\\
&+ \tau_{m,m_0}(1 - [w_+(m) + w_-(m)]\delta t) + \mathcal{O}(\delta t^2).
\end{align}

This expression can be simplified by subtracting $\tau_{m,m_0}$ from both sides, and dividing by $\delta t$:
\begin{align}
\label{eq:recursion}
w_+(m)\Delta \tau_{m+\Delta m} - w_-(m)\Delta \tau_m + \mathcal{O}(\delta t) = -1.
\end{align}
where we have defined $\Delta \tau_m \equiv \tau_{m,m_0} - \tau_{m-\Delta m, m_0}$. Now we can take the limit $\delta t \to 0$ and drop the $\mathcal{O}(\delta t)$ term, obtaining a recursion relation for $\Delta \tau_m$. Since $w_+(1) = 0$ in Equation (\ref{eq:wp}) above, we have $\Delta \tau_1 = 1/w_-(1)$, and can use Equation (\ref{eq:recursion}) to construct all the $\Delta \tau_m$'s from this starting point. Each iteration multiplies the previous increment $\Delta \tau_{m+\Delta m}$ by $w_+(m)/w_-(m)$, and adds a new term $1/w_-(m)$. Thus we find
\begin{align}
\Delta \tau_m &= \frac{1}{w_-(m)} \nonumber\\
&\,\,\,\,\,\,\,\,\,\,\,\,+ \sum_{m' = m}^1 \frac{w_+(m)w_+(m+\Delta m)\dots w_+(m')}{w_-(m)w_-(m+\Delta m)\dots w_-(m'+\Delta m)}\nonumber\\
\label{eq:compact}
&= \sum_{m' = m}^1 \frac{w_+(m)w_+(m+\Delta m)\dots w_+(m'-\Delta m)}{w_-(m)w_-(m+\Delta m)\dots w_-(m')}
\end{align} 
where the sum is over all the discrete values of the occupancy from $m$ to 1 in increments of $\Delta m$. The compact form in the second line is obtained by understanding the $m' = m$ term to have a numerator equal to 1. 

We can rewrite the terms of this series in a more transparent form using Equation (\ref{eq:detailedw}) above, which relates ratios of rates to ratios of steady-state probabilities:
\begin{align}
\frac{w_+(m)}{w_-(m+\Delta m)} = \frac{p_{\rm ss}(m+\Delta m)}{p_{\rm ss}(m)}.
\end{align}
Substituting this expression into Equation (\ref{eq:compact}), we obtain:
\begin{align}
\Delta \tau_m &= \sum_{m' = m}^1 \frac{p_{\rm ss}(m+\Delta m)p_{\rm ss}(m+2\Delta m)\dots p_{\rm ss}(m')}{w_-(m)p_{\rm ss}(m)p_{\rm ss}(m+\Delta m)\dots p_{\rm ss}(m'-\Delta m)}\nonumber\\
&= \sum_{m' = m}^1 \frac{p_{\rm ss}(m')}{w_-(m)p_{\rm ss}(m)}
\end{align}
where we have canceled out all terms that occur in both the numerator and denominator. Note that the $m' = m$ term behaves correctly, since $p_{\rm ss}(m')$ will cancel the $p_{\rm ss}(m)$ from the denominator, leaving the expected value $1/w_-(m)$.

The final step to obtain $\tau_{m^*,m_0}$ is to note the trivial fact that $\tau_{m_0,m_0} = 0$, since it takes no time to reach $m_0$ starting from $m_0$. This lets us construct $\tau_{m^*,m_0}$ by adding up $\Delta\tau_m$'s from $m_0$ to $m^*$:
\begin{align}
\tau_{\rm diss} \equiv \tau_{m^*,m_0} &= \sum_{m = m_0}^{m^*} \Delta \tau_m\nonumber\\
&= \sum_{m = m_0}^{m^*} \sum_{m' = m}^1 \frac{p_{\rm ss}(m')}{w_-(m)p_{\rm ss}(m)}.
\end{align}
This is the expression used in Equation (10) of the main text.

In the large $N$ limit, we can recast this in a more intuitive form by replacing $p_{\rm ss}(m)$ with the asymptotic expression $e^{-Nf(m)}$ (noting that the normalization constant cancels out). This yields
\begin{align}
\tau_{\rm diss} &\approx N^2 \int_{m_0}^{m^*} dm \int_{m}^1 dm'\, e^{N[f(m)-f(m')]}\frac{1}{w_-(m)}.
\end{align}
As $N\to\infty$, this integral can be evaluated via Laplace's method by finding the point where $f(m)-f(m')$ is maximized. This occurs when $m'$ is the location of the high-$m$ local minimum $m^*$, and $m$ is the location of the local maximum between the two states $m^\dagger$, yielding the asymptotic expansion (cf. \cite[eq. 6.4.35]{Bender1999}):
\begin{align}
\tau_{\rm diss} &\approx e^{N[f(m^\dagger)-f(m^*)]}\frac{2\pi N}{\sqrt{|f''(m^*)f''(m^\dagger)|}} \frac{1}{w_-(m^\dagger)}.
\end{align}
For large $N$, the exponential part dominates the qualitative behavior of $\tau_{\rm diss}$ as the free energy landscape changes, which is the meaning of Equation (11) in the main text.

\subsection{Comparison with Equilibrium Statistics}

\begin{figure*}
	\centering
	\includegraphics[width=17cm]{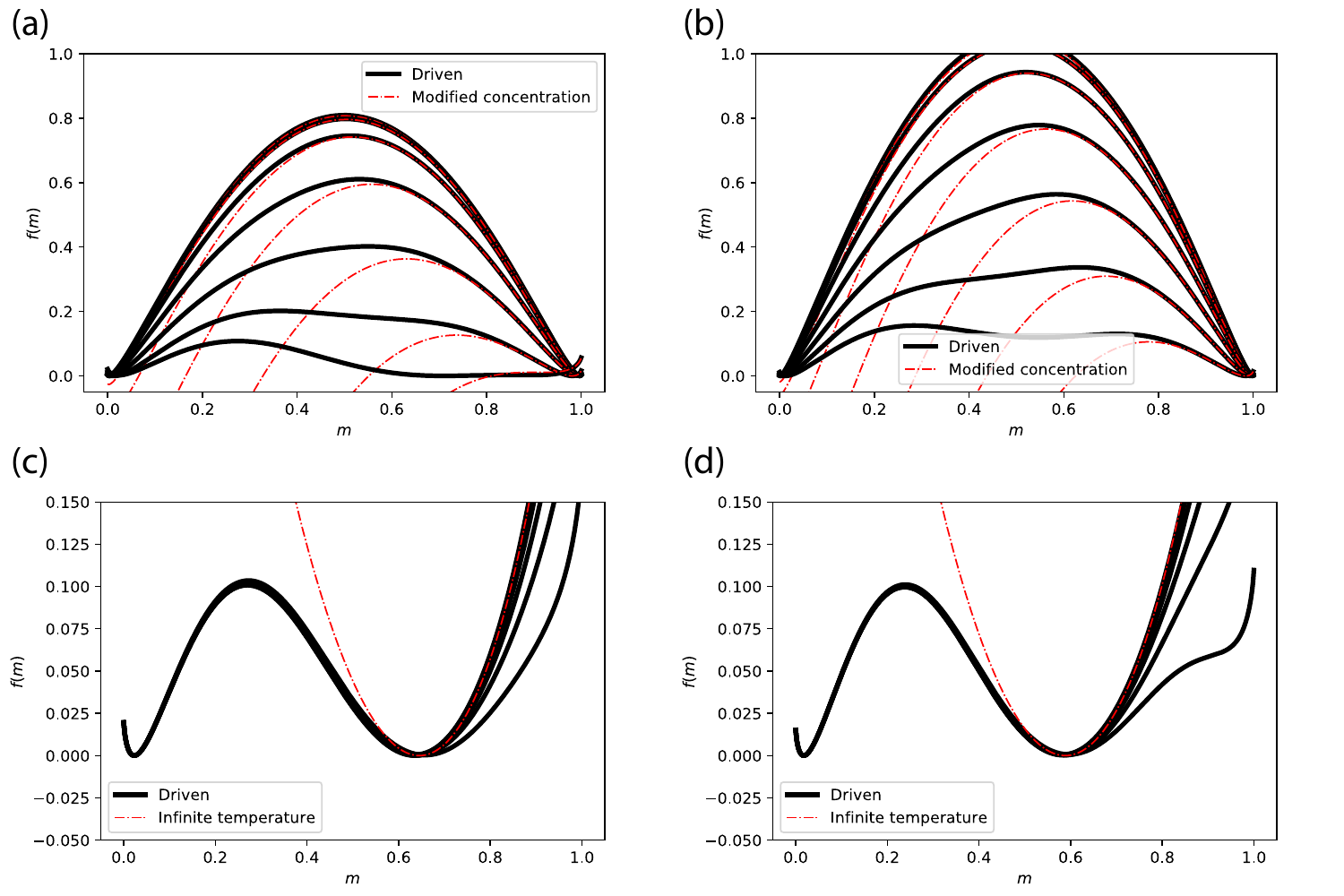}
	\caption{{\bf Equilibrium approximations are generalizable}. Color online. (a) Effective free energy of the driven system and the equilibrium free energy under a reduced monomer concentration, for $\Delta\mu$ up to 10 $k_B T$ and $\beta J = 12$. (b) Same comparison, but with $\beta J = 14$. (c) Effective free energy of the driven system and the equilibrium free energy at infinite temperature, for $\Delta\mu$ above 11 $k_B T$ and $\beta J = 12$. (d) Same comparison, but with $\beta J = 14$.}
	\label{fig:supp}
\end{figure*}

Figure 3 of the main text illustrates how local steady-state fluctuations near the high-$m$ local minimum can be approximated by equilibrium distributions with altered parameters. The example at low $\Delta\mu$ was well approximated by the equilibrium free energy landscape at the same temperature and coupling, but reduced monomer concentration $c_A$. For high $\Delta\mu$, the matching landscape had infinite temperature, with constant off-rate per monomer equal to $k_I = 0.01$. Figure \ref{fig:supp} confirms the validity of these approximations for the whole range of $\Delta\mu$ values displayed in Figure 2a, and for both of the $\beta J$ values from that figure that exhibit a nonequilibrium phase transition. 

The low-$\Delta\mu$ approximation works because the local minimum $m^*$ is only slightly perturbed from its equilibrium value by the active regeneration, and remains within the range of possible values for the high-$m$ local minimum at the original temperature. In the vicinity of $m^*$, the nonequilibrium contribution $f_{\rm neq}(m) \equiv f(m)-f_{\rm eq}(m)$ to the effective free energy landscape can be approximated as a linear function of $m$: $f_{\rm neq}(m) = f_{\rm neq}(m^*) + f_{\rm neq}'(m^*) (m-m^*) + \mathcal{O}[(m-m^*)^2]$. Up to an arbitrary additive constant, the effect of the driving is to add an extra linear term $m f_{\rm neq}'(m^*)$ to $f_{\rm eq}$. If we look at the form of $f_{\rm eq}(m)$ in Equation (\ref{eq:feq}), we see that this is equivalent to changing the monomer concentration to $c_A^{\rm eff} = c_A e^{-f_{\rm neq}'(m^*)}$. 

This argument fails when $\Delta\mu$ crosses the threshold of the nonequilibrium phase transition, and $m^*$ moves outside of the range of values achievable by changing concentration. In this high-$\Delta\mu$ regime, setting the concentration equal to $c_A^{\rm eff} =  c_A e^{-f_{\rm neq}'(m^*)}$ destroys the high-$m$ local minimum in the equilibrium landscape, leaving just a single minimum near $m=0$. 

To understand the fluctuations in this regime, we observe that when $k_A = k_I e^{-\beta (\Delta \mu_0-Jm)} \ll 1$, almost every transition to the inactive state ends in dissociation from the lattice, so the dissociation rate through this pathway is equal to the inactivation rate $k_I$.  If it is also the case that $k_I \gg e^{-\beta Jm}$, then this is the dominant pathway for dissociation, and we can write
\begin{align}
w_-(m) \approx N m k_I.
\end{align}
These same requirements guarantee that 
\begin{align}
w_+(m) \approx N (1-m)c_A
\end{align}
as long as $c_I < c_A$. The $m$-dependence of both of these rates is identical to that of the undriven dynamics in the absence of inter-particle coupling ($\beta J = 0$), but with the dissociation rate reduced from 1 to $k_I$. The arguments of Section \ref{sec:free} above then yield $f(m) \approx m \ln m + (1-m) \ln (1-m) - m \ln c_A/k_I$. If this regime includes the high-$m$ local minimum of $f(m)$, as in Figure 3b, then the steady-state fluctuations will resemble those of the equilibrium system at $\beta J = 0$, and the effective temperature is infinite.

\end{document}